\documentclass{aa}
\usepackage{graphicx}
\usepackage{natbib}
\usepackage[varg]{txfonts}

\newcommand{\ltsima}{$\; \buildrel < \over \sim \;$}
\newcommand{\simlt}{\lower.5ex\hbox{\ltsima}}
\newcommand{\gtsima}{$\; \buildrel > \over \sim \;$}
\newcommand{\simgt}{\lower.5ex\hbox{\gtsima}}

\newcommand{\cgs}{ ${\rm erg~cm}^{-2}~{\rm s}^{-1}$} 
\newcommand{\lum}{\rm erg~s$^{-1}$}

\def\lesssim{\mathrel{\hbox{\rlap{\hbox{\lower4pt\hbox{$\sim$}}}\hbox{$<$}}}}
\def\gtrsim{\mathrel{\hbox{\rlap{\hbox{\lower4pt\hbox{$\sim$}}}\hbox{$>$}}}}

\def\arcsec{\hbox{$^{\prime\prime}$}}
\def\micron{\hbox{$\mu$m}}

\def\ab1450{$AB_{1450(1+z)}$}

\def\xray{\hbox{X-ray}}

\def\oiii{\hbox{[O\ {\sc iii}}]}
\def\oii{\hbox{[O\ {\sc ii}}]}
\def\nev{\hbox{[Ne\ {\sc v}}]}

\def\lsun{\hbox{$L_\odot$}}

\newcommand\phn{\phantom{0}}%
\def\09104{IRAS~09104$+$4109}
\def\I09104{I09104}
\def\fhx{$F_{\rm 2-10\ keV}$}

\def\edd_ratio{$\log\ L_{\rm bol}/L_{\rm Edd}$}

\def\l58{{$(\lambda L_{\lambda})_{\mbox{{\rm \scriptsize 5.8\micron}}}$}}
\def\lmir2{{$(\lambda L_{\lambda})_{\mbox{{\rm \scriptsize 12.3\micron}}}$}}
\def\s1{{S$_{\mbox{{\rm \scriptsize 3.6\micron}}}$}}
\def\irac2{{S$_{\mbox{{\rm \scriptsize 4.5\micron}}}$}}
\def\f3{{S$_{\mbox{{\rm \scriptsize 5.8\micron}}}$}}
\def\mic8{{S$_{\mbox{{\rm \scriptsize 8\micron}}}$}}
\def\f24{{F$_{\mbox{{\rm \scriptsize 24\micron}}}$}}

\def\chandra{{\it Chandra\/}}

\def\heao1{{\it HEAO-1\/}}

\def\nustar{{\it NuSTAR\/}}

\def\xmm{{XMM-{\it Newton\/}}}

\def\swift{{\it Swift\/}}
\def\integral{{\it Integral\/}}
\def\nature{{Nature}}
\def\science{{Science}}
\begin{document} 
\title{The space density of Compton-thick AGN at $z$$\approx$0.8 in the 
zCOSMOS-Bright Survey}

\author{
C. Vignali\inst{1,2}
\and
M. Mignoli\inst{2}
\and
R.~Gilli\inst{2}
\and
A.~Comastri\inst{2}
\and
K.~Iwasawa\inst{3}
\and
G.~Zamorani\inst{2}
\and
V.~Mainieri\inst{4}
\and  
A.~Bongiorno\inst{5}
}

\institute{
Dipartimento di Fisica e Astronomia, Universit\`a degli Studi di Bologna, 
Viale Berti Pichat 6/2, 40127 Bologna, Italy
\email{cristian.vignali@unibo.it}
\and
INAF -- Osservatorio Astronomico di Bologna, Via Ranzani 1, 40127 Bologna, 
Italy
\and
ICREA and Institut de Ci\`encies del Cosmos (ICC), Universitat de Barcelona (IEEC-UB), Mart\'i i Franqu\`es 1, 08028, Barcelona, Spain
\and 
European Southern Observatory, Karl-Schwarzschild-Straße 2, 85748, Garching, Germany
\and 
INAF -- Osservatorio Astronomico di Roma, Via Frascati 33, 00040 Monteporzio 
Catone, Roma, Italy
}

\date{Received 11 August 2014; Accepted 17 September 2014}


\abstract
{The obscured accretion phase in black hole growth is a key ingredient in 
many models linking the Active Galactic Nuclei (AGN) activity with the 
evolution of their host galaxy. 
At present, a complete census of obscured AGN is still missing, 
although several attempts in this direction have been carried out recently, 
mostly in the hard X-rays and at mid-infrared wavelengths.}
{The purpose of this work is to assess the reliability of the \nev\ emission 
line at 3426\AA\ to pick up obscured AGN up to z$\approx$1 by assuming that 
\nev\ is a reliable proxy of the intrinsic AGN luminosity and 
using moderately deep \xray\ data to characterize the amount of obscuration.}
{A sample of 69 narrow-line (Type~2) AGN at $z$$\approx$0.65--1.20 were 
selected from the 20k-zCOSMOS Bright galaxy sample on the basis of the 
presence of the \nev3426\AA\ emission. The \xray\ properties of these galaxies 
were then derived using the \chandra-COSMOS coverage of the field; 
the \xray-to-\nev\ flux ratio, coupled with \xray\ spectral and stacking 
analyses, was then used to infer whether Compton-thin or Compton-thick 
absorption were present in these sources. Then the \nev\ luminosity function 
was computed to estimate the space density of Compton-thick AGN at 
$z\approx0.8$.}
{Twenty-three sources were detected by \chandra, and their properties are 
consistent with moderate obscuration (on average, 
$\approx$ a few $\times10^{22}$~cm$^{-2}$). 
The \xray\ properties of the remaining 46 \xray\ undetected Type~2 AGN 
(among which we expect to find the most heavily obscured objects) were 
derived using \xray\ stacking analysis. Current data, supported by Monte-Carlo 
simulations, indicate that a fraction as high as $\approx$~40\% of the present 
sample is likely to be Compton thick. 
The space density of Compton-thick AGN with log$L_{2-10 keV}>43.5$ at $z=0.83$ 
is $\Phi_{Thick} = (9.1\pm2.1) \times 10^{-6} $ Mpc$^{-3}$, in good agreement 
with both X-ray background model expectations and the previously measured 
space density for objects in a similar redshift and luminosity range. 
We regard our selection technique for Compton-thick AGN as clean but 
not complete, since even a mild extinction in the narrow-line region can 
suppress \nev\ emission. Therefore, our estimate of their space density 
should be considered as a lower limit.}
{}

\authorrunning{C. Vignali et al.}
\titlerunning{Optically selected obscured AGN at $z\approx0.8$}

\maketitle


\section{Introduction}
The existence of a close link between super massive black holes (SMBHs) and 
their host galaxies has found support in many works over the last twenty years, 
starting from the seminal paper by \cite{kormendy1995}. 
Several relations in the local Universe between the mass of the black holes 
and the properties of the host galaxies (e.g., their velocity dispersion; 
\citealt{gebhardt2000}; \citealt{ferrarese2000}) are signatures of 
this connection. 
Most active galactic nuclei (AGN)-galaxy co-evolution models predict 
the existence of a dust-enshrouded phase characterized by rapid 
SMBH growth and active star formation, largely triggered by 
multiple galaxy mergers and encounters (e.g., 
\citealt{silk1998}; \citealt{dimatteo2005}; \citealt{menci2008}; 
\citealt{zubovas2012}; \citealt{lamastra2013}). This phase is associated 
to obscured AGN growth in strongly star-forming galaxies, 
as already shown a decade ago by \cite{alexander2005} using sub-millimeter 
selection and deep \xray\ data. 
Then massive AGN-driven outflows 
blow away most of the cold gas reservoir; the AGN become unobscured (i.e., 
Type~1) and, when the phase of significant nuclear accretion is over due to 
the limited amount of remaining available gas, a population of 
``red-and-dead'' gas-poor elliptical galaxies is what is left 
(e.g., \citealt{hopkins2008}; \citealt{cattaneo2009}). 
If, on the one hand, this scenario seems to provide a valid explanation 
for the most luminous and massive systems, for which a large fraction of 
the mass is assembled in relatively short periods ($\approx10-100$~Myr) of 
intense nuclear and star-forming activity, on the other hand the bulk of 
galaxies and SMBHs are likely to grow their mass in a secular 
(i.e., ``smooth'') mode over timescales of Gyrs (e.g., \citealt{daddi2007a}; 
\citealt{hickox2009}; \citealt{villforth2014}, and references therein).

Besides the many observational evidences of feedback processes in terms 
of outflows in molecular and neutral/ionized gas, at both low and high redshift 
(e.g., \citealt{nesvadba2008}; \citealt{feruglio2010}; \citealt{maiolino2012}; 
\citealt{harrison2012,harrison2014}; \citealt{cicone2014}; 
\citealt{brusa2014}), including signatures in the 
\xray\ band (e.g., \citealt{chartas2002}; \citealt{tombesi2012}, 
and references therein), this ``picture'' still needs to be confirmed in most 
of its aspects, in particular for what concerns the key phase during which 
large amounts of gas are funneled to the center, thus 
inducing both obscured accretion and star formation (e.g., 
\citealt{treister2010}). Type~2 (obscured) AGN are witnesses and, 
at the same time,  main actors of such phase; 
finding them across cosmic time is therefore 
crucial to place constraints on AGN vs. galaxy co-evolution models (see, e.g., 
the reviews by \citealt{merloni2013} and \citealt{vignali2014}) 
as well as on \xray\ background synthesis models (XRB; e.g., 
\citealt{gilli2007}; \citealt{ballantyne2009}; \citealt{treister2009}; 
\citealt{akylas2012}; \citealt{shi2013}; see also \citealt{moretti2012}). 

A complete census of absorbed AGN, especially of the most heavily 
obscured (called Compton thick, having column densities above 
$1.5\times10^{24}$~cm$^{-2}$; see \citealt{comastri2004} for 
a review) up to high redshift cannot be achieved using either a single 
selection or a single observing-band method. 
%
Although the hard X-ray band up to $\approx$100--150~keV 
is potentially appropriate to provide an almost unbiased census 
of obscured AGN, since both Compton-thin 
(with $N_{H}<1.5\times10^{24}$~cm$^{-2}$) and Compton-thick AGN 
(at least the sources having column densities up to $\approx10^{25}$~cm$^{-2}$) 
can be detected by hard \xray\ instruments, the sensitivity of 
\swift/BAT and \integral/IBIS surveys ($\approx10^{-11}$~\cgs; e.g., 
\citealt{tueller2008}; \citealt{beckmann2009}; \citealt{burlon2011}; 
\citealt{ajello2012}; \citealt{vasudevan2013}) is such to  
limit this kind of investigations to the local Universe. Interesting 
prospects might come from the much higher sensitivity 
(a factor of $\approx$100) hard \xray\ imaging instrument onboard \nustar\ 
(e.g., \citealt{alexander2013}; \citealt{lansbury2014}). 

A proper evaluation of obscured AGN at higher redshifts and at fainter \xray\ 
fluxes requires deep \xray\ surveys with \chandra\ and \xmm\ (e.g., 
\citealt{tozzi2006}; \citealt{georgantopoulos2009,georgantopoulos2013}; 
\citealt{comastri2011}; \citealt{feruglio2011}; \citealt{gilli2011, gilli2014}; 
\citealt{iwasawa2012a}; \citealt{brightman2012}; \citealt{vito2013}; 
\citealt{buchner2014}; \citealt{brightman2014}). 
However, even the deepest \xray\ exposures currently available miss a 
significant number of very obscured AGN, 
especially the most heavily Compton-thick AGN at low-medium redshift 
(because of the energy bandpass, poor counting statistics and sensitivity), 
hence a not negligible fraction 
of the accretion power in the Universe, and convey the idea that 
a multi-wavelength approach to the search for obscured AGN is mandatory. 

The availability of deep radio, optical and mid-infrared (mid-IR) data has 
recently allowed the definition of complementary methods to pick up obscured 
AGN up to high redshifts. 
In particular, sources with weak emission in the optical band (due to 
dust extinction) and relatively bright mid-IR emission (due to the 
disc emission being thermally reprocessed by the AGN torus) 
can be considered good obscured AGN candidates 
(e.g., \citealt{sansigre2005}; \citealt{houck2005}; 
\citealt{weedman2006}; \citealt{dey2008}), unless a significant contribution 
to the mid-IR comes from star-formation processes (PAH features and 
continuum emission). This selection of obscured AGN, albeit not complete, 
represents a step forward with respect to mid-IR colour-colour diagrams, where 
separating heavily obscured AGN from the remaining AGN populations 
is not trivial (e.g., \citealt{lacy2004,lacy2013}; \citealt{stern2005}; 
\citealt{donley2012}; \citealt{castello_mor2013}; \citealt{mateos2013}).
However, the selection of obscured AGN by means of long-wavelength data needs 
to be corroborated by \xray\ information, as plenty of works have shown in the 
last decade (e.g., \citealt{polletta2006}; \citealt{fiore2008}; 
\citealt{lanzuisi2009}; \citealt{severgnini2012}; \citealt{delmoro2013}). 
This observational approach has led many authors to place constraints, 
for the first time, to the space density of Compton-thick AGN at high redshifts 
($z\approx2-3$; \citealt{daddi2007b}; \citealt{fiore2009}; 
\citealt{bauer2010}; \citealt{alexander2011}). 

Obscured AGN can be selected also on the basis of narrow, high-ionization 
emission lines, like \oiii5007\AA\ and \nev3426\AA. 
These lines are produced in the narrow-line region (NLR), so they do not suffer 
from the effects of attenuation ascribed to the torus obscuring material 
(although some extinction within the NLR may be present).  
As such, these lines can be considered good proxies of the nuclear emission 
from the AGN. Previous works have shown that coupling AGN selection in one of 
these lines with \xray\ observations may be effective in finding highly 
reliable Compton-thick AGN candidates (e.g., \citealt{vignali2006,vignali2010}; 
\citealt{ptak2006}; \citealt{lamassa2009}; \citealt{lamastra2009}; 
\citealt{gilli2010}, G10 hereafter; \citealt{jia2013}; 
but see also \citealt{trouille2010}). Once again, \xray\ 
information is crucial to infer the nature of such sources, at least on a 
statistical ground, with further enlightening indications coming from the 
mid-IR emission (e.g., \citealt{vignali2010}; \citealt{goulding2011}; 
see also \citealt{gandhi2009}). 

In this paper we present the \xray\ properties of a sample of Type~2 AGN 
selected from the 20k-zCOSMOS Bright survey (\citealt{mignoli2013}, 
M13 hereafter). 
The optical selection and \xray\ (mostly \chandra) coverage are shown 
in Sect.~\ref{sample_and_xray}. 
Results from \xray\ spectroscopy and stacking analyses are reported in 
Sect.~\ref{xray_results}, along with the estimated fraction of Compton-thick 
AGN. Their space density at $z\approx0.8$ is then computed in 
Sect.~\ref{space_density_obscured}, where current results are also compared 
with expectations based on AGN synthesis models of the XRB. 
Conclusions are then reported in Sect.~\ref{conclusions}. 

Hereafter we adopt a cosmology with $H_{0}$=70~km~s$^{-1}$~Mpc$^{-1}$, 
$\Omega_{\rm M}$=0.3 and $\Omega_{\Lambda}$=0.7.

\section{The sample: optical selection and \textit{Chandra} coverage}
\label{sample_and_xray}
The selection of the \nev\ Type~2 AGN sample has been extensively 
described and discussed in M13. In the following, we will report on the main 
issues related to the optical selection process, the \xray\ coverage, the 
fraction of \xray\ detections, and the \xray\ flux estimates needed to 
place sources in the X/NeV flux ratio diagnostic histogram (G10). 

Type~2 AGN were selected from the zCOSMOS spectroscopic survey 
\citep{lilly2007,lilly2009}, which was designed to observe 
$\approx$20,000 galaxies at I-band 
magnitudes brighter than 22.5 over the entire $\approx$~2~square~degree 
Cosmic Evolution Survey (COSMOS) field \citep{scoville2007}, at a medium 
resolution (R$\approx$600) and in the 5500--9700\AA\ spectral range 
with the VLT instrument VIMOS. Hereafter, this spectroscopic survey will be 
referred to as the 20k-zCOSMOS Bright survey. 

Ninety-four Type~2 AGN were selected from the 20k-zCOSMOS survey in 
the redshift range $\approx$0.65--1.20. The lower redshift boundary guarantees 
that both \nev3346\AA\ and \nev3426\AA\ emission lines fall within the spectral 
coverage; the choice of the upper boundary is due to the strong fringing of 
VIMOS spectra at long wavelengths, which makes the process of line detection 
more difficult and less reliable. 
The Type~2 classification was based on accurate analysis of all of the 
optical spectra, showing only narrow features. 
The presence of significant emission from broad emission-line components is 
ruled out, as clearly shown in the composite spectrum of the Type~2 AGN 
sample in M13 (see their Fig.~2). 
The rest-frame equivalent width (EW) of the \nev\ emission line spans  
the range $\approx2-35$\AA, with a median value of $\approx8$\AA. 

Seventy-one of these \nev-selected Type~2 AGN fall in the \chandra-COSMOS 
mosaic ($\approx$~0.9~square~degrees). Two of these sources fall in a bad 
position of the mosaic and were therefore discarded. 
The choice of focusing on \chandra\ instead of \xmm\ 
to derive the \xray\ properties of this sample is due to the 
typically deeper exposure in the \chandra\ mosaic ($\approx$160~ks over the 
inner 0.45~square~degrees -- thus reaching fainter flux limits than XMM-COSMOS 
by a factor of $\approx$3.5 -- and an outer region of 
$\approx$0.4~square~degrees with an effective exposure of up to $\approx$80~ks; 
see \citealt{puccetti2009, elvis2009, civano2012}, C12 hereafter), 
and to its better performances when investigation concerns \xray\ faint 
sources. The low and ``smooth'' background level of \chandra\ observations 
is also a key ingredient for stacking analyses. 

Twenty-three \nev-selected Type~2 AGN were detected by \chandra\ 
(see Table~\ref{neon23}) within 1.2\arcsec\ from the optical position. 
The median displacement between the \xray\ and the optical counterpart, 
0.46\arcsec, and the maximum likelihood technique presented by C12 assures 
the reliability of the associations. 
We have visually inspected the four sources 
with a counterpart at a distance larger than 1 arcsec and found they are 
good matches. 

\begin{figure}
\centering
\includegraphics[width=\hsize]{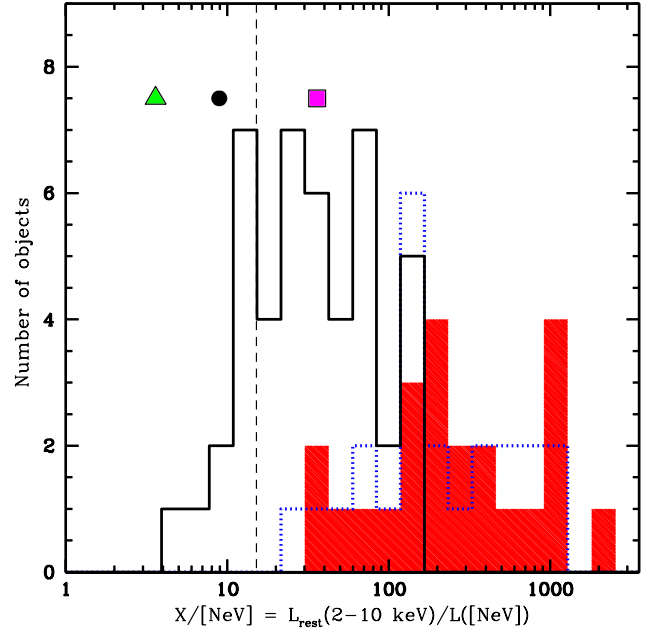}
\caption{Distribution of the rest-frame 2--10~keV (with no correction for 
absorption) to [NeV] luminosity ratio (X/NeV) of the 69 zCOSMOS Type~2 AGN 
with reliable \xray\ photometry. 
\xray\ detections are reported as either red (filled) or blue (dotted) 
histograms; 
in particular, the red histogram refers to \xray\ photometry obtained from the 
{\sc Acis Extract} analysis carried out in the rest-frame 2--10~keV band, 
while the blue histogram reports the \xray\ flux values derived from direct 
\xray\ spectral fitting in the same energy band (see text for details). 
The black histogram refers to the \xray\ undetected \nev-selected Type~2 AGN. 
The vertical dashed line shows the threshold defined by G10 for Compton-thick 
(leftward direction) and Compton-thin (rightward direction) sources. 
All of the \xray\ detected sources are likely Compton thin using the X/NeV 
threshold of 15 adopted by G10. 
The filled symbols represent the average X/NeV value obtained from the \xray\ 
stacking analysis (black circle: stack of the 46 \xray\ undetected Type~2 AGN; 
green triangle: stack of the 22 sources with ``nominal'' upper limit on 
the X/NeV ratio $<$30; magenta square: stack of the 24 sources with 
``nominal'' upper limit on the X/NeV ratio $>$30; see Sect.~\ref{section46} 
for details). 
Their y-axis position is arbitrary.}
\label{xne_histo}
\end{figure}

To characterize the sources in terms of obscuration, we used the 
X/NeV flux ratio diagram (as in G10), where $X$ corresponds to the rest-frame 
2--10~keV flux with no absorption correction, 
and $NeV$ is 
the flux of the emission line derived from the zCOSMOS spectra after 
applying aperture corrections, whose median value corresponds to a factor of 
$\approx1.5$ (see $\S$6 of M13). 
The rest-frame 2--10~keV band corresponds to the observed-frame 
$\approx$1--5.5~keV band. 
We have extracted the 23 [NeV]-selected source information 
from the multiple pointings of the \chandra-COSMOS mosaic using the procedure 
outlined in M13 and derived \xray\ photometry in the chosen energy interval. 
Due to the tiles in the C-COSMOS field, every source can be observed in more 
than one pointing (up to six) and at different locations within the ACIS-I 
field-of-view, hence the source count distribution is characterized by 
different (i.e., multiple) point spread function (PSF) sizes and shapes. 
To properly account for these effects, we have used the {\sc Acis Extract} 
software \citep{broos2010}, which extracts the source 
counts from each observation using the 90\% of the encircled energy fraction 
at 1.5~keV at the source position and then corrects for the chosen aperture. 
At this stage, counts were extracted in the rest-frame 2--10~keV for each 
source and divided by the effective (i.e., ``vignetting-corrected'') exposure 
time at each source position (averaged over an area of 5-arcsec radius to 
minimize the possible effects of small-scale fluctuations in the timemaps) 
to obtain a count rate. The \xray\ flux has then 
been obtained by a simple count rate to flux conversion, assuming a powerlaw 
model with photon index $\Gamma$=1.4 modified by Galactic absorption. 
The corresponding X/NeV flux ratios are plotted in 
Fig.~\ref{xne_histo} (red filled region). 

Along with photometry, \xray\ spectra were also extracted and fitted with a 
simple model (a powerlaw modified by Galactic absorption), due to the limited 
photon statistics for most of these sources (see $\S$\ref{section23}). 
This \xray\ spectral analysis allowed us to derive a further estimate for 
the rest-frame 2--10~keV fluxes (see Table~\ref{neon23}, where two 
X/NeV columns are shown, corresponding to the two methods used to derive 
\xray\ fluxes). The X/NeV flux ratios corresponding to this basic \xray\ 
spectral analysis are plotted in Fig.~\ref{xne_histo} as a blue dotted 
histogram. Details on the \xray\ properties of these 23 Type~2 AGN 
are presented in Sect.~\ref{section23}. 

A similar, {\sc Acis Extract}-based procedure has been adopted for the 
\xray\ photometry of the remaining 46 \nev-selected Type~2 AGN with 
no \xray\ detection in the C12 catalog 
(and no clear \xray\ emission in the 0.5--7~keV images). 
To derive \xray\ flux upper limits, at each source optical position counts 
were extracted taking into account the PSF size and shape in the observed 
band equivalent to the rest-frame 2--10~keV energy interval. One-sigma count 
upper limits have then been converted into count rates using the exposures 
derived from the timemaps at each source position, and count rates 
were then converted into \xray\ flux upper limits assuming a powerlaw with 
photon index $\Gamma$=$-$0.4, which is broadly consistent with a 
reflection-dominated spectrum. These values were then used to derive the 
X/NeV flux ratio upper limits reported in Table~\ref{neon46}, similarly to 
the analysis presented in M13. We note that Monte-Carlo runs were 
then used to properly interpret the results obtained for the \xray\ 
undetected Type~2 AGN given their limited photon statistics 
(see Sect.~\ref{section46} for details).

The different behaviour in the \xray\ band between the sample of 
23 \xray\ detected and that of 46 \xray\ undetected Type~2 AGN 
cannot be ascribed to a difference in the exposure time; 
the former sample has an average exposure of $\approx$120~ks, 
while the latter is characterized by a slightly larger exposure, 
$\approx$128~ks. Furthermore, the two samples have comparable average redshift 
(z=0.85 vs. z=0.84) and similar \nev\ flux ($3.1\times10^{-17}$ vs. 
$2.8\times10^{-17}$~\cgs). 
Therefore, we motivate our choice of a different photon index for 
the count rate to flux conversion for the two samples under the hypothesis 
that \xray\ undetected Type~2 AGN may suffer from heavier absorption 
than \xray\ detected Type~2 AGN. 
%

As a final remark, we note that the present sample of Type~2 AGN is far from 
being complete, as extensively discussed by M13. In particular, there are 
further $\approx$160 zCOSMOS galaxies in the same redshift interval having 
the properties of Type~2 AGN (either from optical spectroscopy or \xray\ 
analysis or both) without the presence of \nev\ emission. The 
mean optical extinction of this sample (derived from the H$\beta$/H$\gamma$ 
flux ratio, \hbox{$\langle$E(B$-$V)$\rangle\,$=$\,0.27$}) is higher than that 
derived for the \nev-selected Type~2 AGN discussed in this paper 
(\hbox{$\langle$E(B$-$V)$\rangle\,$=$\,0.18$}); the higher extinction is likely 
causing the lack of appreciable \nev\ emission in the ``additional'' sample of 
Type~2 AGN (see Fig.~6 of M13). 
Summarizing, the AGN selection based on \nev\ emission can be considered 
clean (because of the ionization potential of this line, unambiguously 
ascribed to accretion power) but not complete. 

\section{X-ray results}
\label{xray_results}

\begin{table*}
\caption{List of X-ray detected [NeV]-selected Type~2 AGN}
\label{neon23}
\centering
\footnotesize
\begin{tabular}{cccccccccccr}
\hline
  \multicolumn{1}{c}{zCOSMOS\_ID} &
  \multicolumn{1}{c}{RA$_{\rm \small (J2000)}$} &
  \multicolumn{1}{c}{DEC$_{\rm \small (J2000)}$} &
  \multicolumn{1}{c}{z} &
  \multicolumn{1}{c}{I mag} &
  \multicolumn{1}{c}{CID} &
  \multicolumn{1}{c}{T$_{\rm Expo}$} &
  \multicolumn{1}{c}{X/[NeV]} &
  \multicolumn{1}{c}{Xs/[NeV]} &
  \multicolumn{1}{c}{Ncts} &
  \multicolumn{1}{c}{\fhx} &
  \multicolumn{1}{c}{XMMID} \\
\hline
  \multicolumn{1}{c}{(1)} &
  \multicolumn{1}{c}{(2)} &
  \multicolumn{1}{c}{(3)} &
  \multicolumn{1}{c}{(4)} &
  \multicolumn{1}{c}{(5)} &
  \multicolumn{1}{c}{(6)} &
  \multicolumn{1}{c}{(7)} &
  \multicolumn{1}{c}{(8)} &
  \multicolumn{1}{c}{(9)} &
  \multicolumn{1}{c}{(10)} &
  \multicolumn{1}{c}{(11)} &
  \multicolumn{1}{c}{(12)} \\
\hline\hline
810378  & 150.467621 & 1.935703 & 0.9707 & 21.81 & {\phn}401 & {\phn}75.99  & {\phn}182.5      & {\phn}181.7            & {\phn}29         & 2.0   & 60205\\
812111  & 150.097778 & 1.845246 & 0.7106 & 21.18 & {\phn}254 & 146.23       & {\phn}162.1      & {\phn}136.7            & {\phn}55         & 0.5   & $-$99\\
813250  & 149.830582 & 1.902132 & 0.7302 & 21.62 & 1019      & 156.12       & {\phn}{\phn}91.4 & {\phn}{\phn}38.5       & {\phn}27         & 0.8   & $-$99\\
813287  & 149.820557 & 1.811719 & 0.7488 & 20.41 & {\phn}221 & {\phn}63.70  & 1057.8           & {\phn}428.3            & {\phn}61         & 0.6   & 5139\\
816439  & 150.510666 & 2.029207 & 0.8988 & 20.92 & {\phn}496 & {\phn}72.13  & {\phn}181.8      & {\phn}146.2            & {\phn}90         & 1.6   & 2387\\
817002  & 150.418167 & 1.976701 & 0.8649 & 20.73 & {\phn}381 & {\phn}91.29  & {\phn}346.3      & {\phn}295.0            & {\phn}77         & 2.1   & 2473\\
817871  & 150.248688 & 1.972335 & 0.6739 & 20.16 & 1508      & 159.31       & {\phn}{\phn}37.6 & {\phn}{\phn}62.2       & {\phn}12         & 0.5   & $-$99\\
817977  & 150.222534 & 2.011670 & 0.9591 & 21.32 & 1169      & 214.06       & {\phn}262.8      & {\phn}135.2            & {\phn}30         & 0.2   & $-$99\\
819469  & 149.893555 & 2.107714 & 0.6865 & 21.35 & {\phn}339 & 161.58       & {\phn}998.0      & {\phn}785.8            & 367              & 2.0   & 64\\
820695  & 149.580399 & 1.967788 & 0.7642 & 21.64 & 1706      & {\phn}36.88  & {\phn}{\phn}69.9 & \dotfill$^{a}$         & {\phn}{\phn}5    & \dotfill  & $-$99\\
820742  & 149.570389 & 1.990572 & 0.9997 & 21.98 & {\phn}173 & {\phn}53.45  & 2079.8           & 1007.1                 & {\phn}79         & 1.5   & 5288\\
825282  & 150.018860 & 2.147779 & 0.9588 & 20.90 & 1126      & 164.75       & {\phn}211.5      & {\phn}171.7            & {\phn}91         & 1.4   & 54534\\
825838  & 149.907333 & 2.169099 & 0.7864 & 21.23 & 1130      & 124.24       & {\phn}252.5      & {\phn}158.2            & {\phn}43         & 0.3   & 497\\
826095  & 149.856766 & 2.273134 & 0.7640 & 20.62 & 2454      & 150.51       & {\phn}{\phn}43.3 & \dotfill$^{a}$         & {\phn}10         & \dotfill  & 364\\ 
829955  & 150.444427 & 2.369805 & 0.8913 & 21.41 & {\phn}717 & 162.04       & {\phn}149.3      & {\phn}108.9            & {\phn}78         & 0.5   & 172\\
831966  & 150.062149 & 2.455000 & 0.7295 & 20.70 & {\phn}110 & 156.57       & {\phn}976.1      & {\phn}747.3            & 309              & 1.8   & 99\\
832900  & 149.884171 & 2.338150 & 1.0230 & 21.30 & {\phn}456 & 181.28       & {\phn}136.9      & {\phn}122.8            & 165              & 2.7   & 413\\
833208  & 149.826523 & 2.396749 & 0.9106 & 22.38 & {\phn}503 & 132.04       & {\phn}637.8      & {\phn}538.4            & {\phn}65         & 1.5   & 5224\\
833510  & 149.768112 & 2.431331 & 0.9444 & 20.94 & {\phn}522 & {\phn}78.00  & {\phn}661.6      & {\phn}565.2            & 160              & 3.3   & 2237\\
833904  & 149.691101 & 2.335542 & 0.8643 & 21.67 & {\phn}426 & {\phn}73.95  & {\phn}354.4      & {\phn}349.0            &  {\phn}43        & 1.3   & 229\\
837988  & 150.274323 & 2.511393 & 0.7034 & 20.72 & {\phn}138 & {\phn}73.65  & 1084.2           & {\phn}947.7            &  {\phn}84        & 1.5   & 158\\
840085  & 149.910660 & 2.554670 & 0.7535 & 20.01 & 1230      & 163.37       & {\phn}{\phn}34.1 & {\phn}{\phn}23.7       &  {\phn}32        & 0.2   & 5008\\
846478  & 150.015457 & 2.665832 & 1.1767 & 22.13 & {\phn}620 & {\phn}79.62  & {\phn}179.8      & {\phn}149.6            &  {\phn}57        & 1.6   & 5427\\
\hline
\end{tabular}
\tablefoot{
(1) Identification number in the 20k-zCOSMOS Bright survey; 
(2) optical right ascension and (3) declination; 
(4) redshift from the 20k-zCOSMOS Bright spectroscopic survey; 
(5) I-band AB magnitude; 
(6) identification number according to the \chandra-COSMOS source catalog 
(C12); 
(7) net (``vignetting-corrected") exposure time in the \chandra\ mosaic (ks); 
(8) X/NeV flux ratio using \xray\ aperture photometry from {\sc Acis Extract} 
to compute the rest-frame 2--10~keV flux (see $\S$\ref{sample_and_xray} for 
details); 
(9) X/NeV flux ratio using the \xray\ flux from spectral fitting of the data 
using {\sc xspec}; 
(10) number of net counts from {\sc Acis Extract} in the observed 
$\approx$0.5--7~keV energy band; 
(11) observed-frame 2--10~keV flux (from \xray\ spectral analysis, in units of 
10$^{-14}$~\cgs); 
(12) identification ID in the XMM-COSMOS source catalog 
(\citealt{brusa2010}; see also \citealt{cappelluti2009}). 
$-99$ indicates that the source was not detected by \xmm. \\
$^{a}$ For this source, the \xray\ photon statistics was not sufficient to 
compute a reliable \xray\ flux in the rest-frame 2--10~keV band.
}
\end{table*}

\subsection{X-ray detected [NeV]-selected Type~2 AGN}
\label{section23}
For the 23 \xray\ detected Type~2 AGN we were able to perform a basic \xray\ 
spectral analysis assuming a powerlaw model. 
The main limitation of this analysis is due to the paucity of 
\xray\ photons available in our data. 
%
The median number of net (i.e., background-subtracted) counts is 60. 
Two sources have 10 or less counts, 
two sources have 140--150 counts (CID=456 and 522, according to C12 source 
nomenclature), and two further sources have 300--350 counts (CID=110 and 339). 

On the one hand, fitting all the \xray\ data together with the same model 
(accounting for the different redshifts and fluxes of the sources) 
would produce biased results, 
given the presence of four AGN with more than 100 counts. 
On the other hand, fitting all the spectra individually using unbinned data and 
{\it C}-statistic (Cash 1979) within {\sc xspec} 
\citep[Version 12.7.0;][]{arnaud1996}
would not constrain properly the \xray\ spectral parameters ($\Gamma$, 
$N_{\rm H}$ and normalization) for most of the 23 sources. 
Therefore, we decided to fit the four sources with 
most counts using $\chi^{2}$ statistics and binned data (10--15 counts per 
bin); the remaining spectra were ``summed'' to produce a rest-frame 1--10~keV 
stacked spectrum (with a spectral resolution of $\approx$0.5~keV) -- 
following the prescriptions extensively described in 
\cite{iwasawa2012a, iwasawa2012b} -- and then fitted (again, adopting the 
$\chi^{2}$ statistics). 

Errors are quoted at the 90\% confidence level for one parameter of interest 
(\hbox{$\Delta\chi^2$=2.71}; \citealt{avni1976}).
Solar abundances (from \citealt{anders_grevesse1989}) and Galactic 
absorption ($N_{\rm H}=1.7\times10^{20}$~cm$^{-2}$; \citealt{kalberla2005}) 
were adopted in all models. 

\begin{figure}
\centering
\includegraphics[angle=0,width=1\hsize]{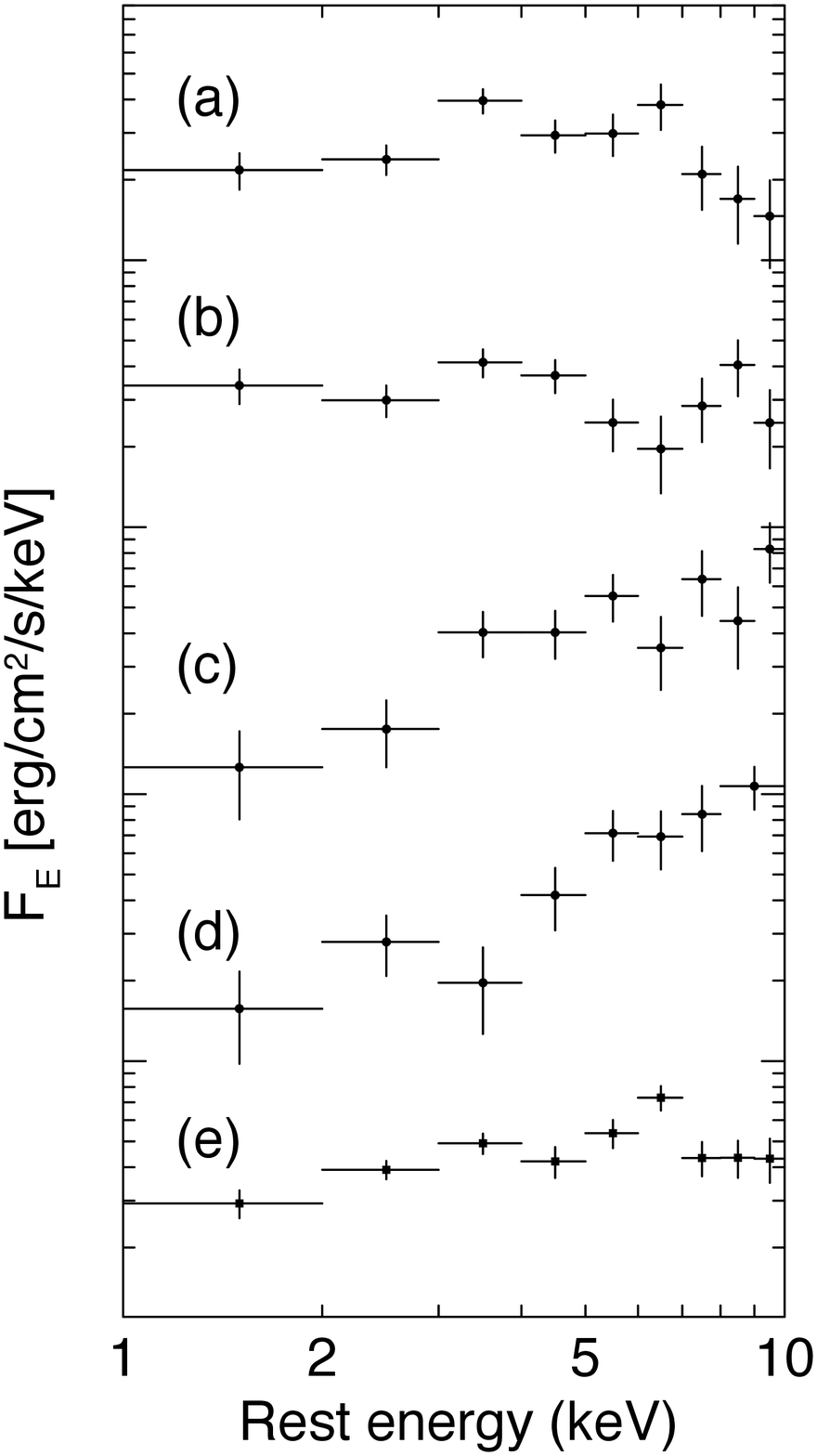}
\caption{Rest-frame \xray\ spectra of the four Type~2 AGN with most 
counts [i.e., above 140 counts; CID339 (a), CID110 (b), CID522 (c), and 
CID456 (d)], characterized by increasing absorption from top to bottom, and 
stacked \xray\ spectrum of the 19 Type~2 AGN with limited photon statistics 
[panel (e)], showing an emission line at an energy consistent with 
iron K$\alpha$. An emission line is also apparent in CID339. 
\xray\ spectra are arbitrarily scaled for visual clarity.}
\label{xrayfitting_n23}
\end{figure}

CID339 and CID110 are characterized by single powerlaws with limited 
absorption at the source redshift 
(N$_{\rm H}=3.8^{+3.2}_{-2.8}\times10^{21}$~cm$^{-2}$ and 
N$_{\rm H}<3.4\times10^{21}$~cm$^{-2}$, respectively, 
once a ``canonical'' AGN photon index $\Gamma$=1.8  -- e.g., 
\cite{piconcelli2005} -- is assumed; see panels (a) and (b) in 
Fig.~\ref{xrayfitting_n23}); 
we note that the best-fitting $\Gamma$ for the two sources 
($\approx$1.6--1.7) are very close to the adopted value. 
An emission line at rest-frame energy $\approx$6.9~keV is tentatively 
detected ($\approx2.5\sigma$) in CID339. 
The other two relatively bright sources, CID522 and CID456, appear more 
absorbed, being characterized by $\Gamma\approx0.9-1.2$ if no absorption is 
included. Fixing the photon index to 1.8 and including obscuration 
in the fit produce a column density of 
$2.7^{+2.3}_{-2.0}\times10^{22}$~cm$^{-2}$ and 
$3.1^{+1.4}_{-1.0}\times10^{23}$~cm$^{-2}$, respectively 
(see panels (c) and (d) in Fig.~\ref{xrayfitting_n23}). 
%
%
The 1--10~keV stacked spectrum of the remaining sources is characterized by a 
flat slope and an emission feature which, at first glance, is consistent 
with iron emission (see the \xray\ stacked spectrum in panel (e) of 
Fig.~\ref{xrayfitting_n23}). 
If no absorption is assumed, $\Gamma=0.8\pm{0.1}$ is derived. Similarly to 
the sources above, the photon index was fixed to 1.8, and the derived column 
density is N$_{\rm H}=2.2^{+0.6}_{-0.5}\times10^{22}$~cm$^{-2}$. 
This column density is consistent with the broad range of X/NeV 
(hence absorption) encompassed by these sources (see the filled red and 
open blue histograms in Fig.~\ref{xne_histo}, and Table~\ref{neon23}). 
The emission-line energy is 6.4$\pm{0.2}$~keV, in agreement with either 
neutral or mildly ionized iron, and its equivalent width is 
EW=0.7$\pm{0.4}$~keV. 
Since a typical signature of strong obscuration is the presence of a strong 
(EW$\gtrsim1$~keV) FeK$\alpha$ emission line on top of a flat \xray\ continuum 
(e.g., \citealt{comastri2011, feruglio2011, iwasawa2012a}), we might 
interpret the EW of the iron line in the stacked spectrum as suggestive of 
the presence of few sources with heavy obscuration. 

The \xray\ spectral results obtained from the stacked spectrum were confirmed 
by fitting the spectra simultaneously (i.e., all of the sources were fitted 
at the same time using the same spectral model, leaving the normalization of 
each source free to vary; here {\it C}-statistic is adopted). 
Assuming $\Gamma=1.8$, we obtain 
N$_{\rm H}=(2.2\pm{0.4})\times10^{22}$~cm$^{-2}$ (consistent with the analysis 
reported above) but the iron line does not emerge as strongly as in the 
stacked spectrum described above. 

\begin{figure}
\centering
%
\includegraphics[width=\hsize]{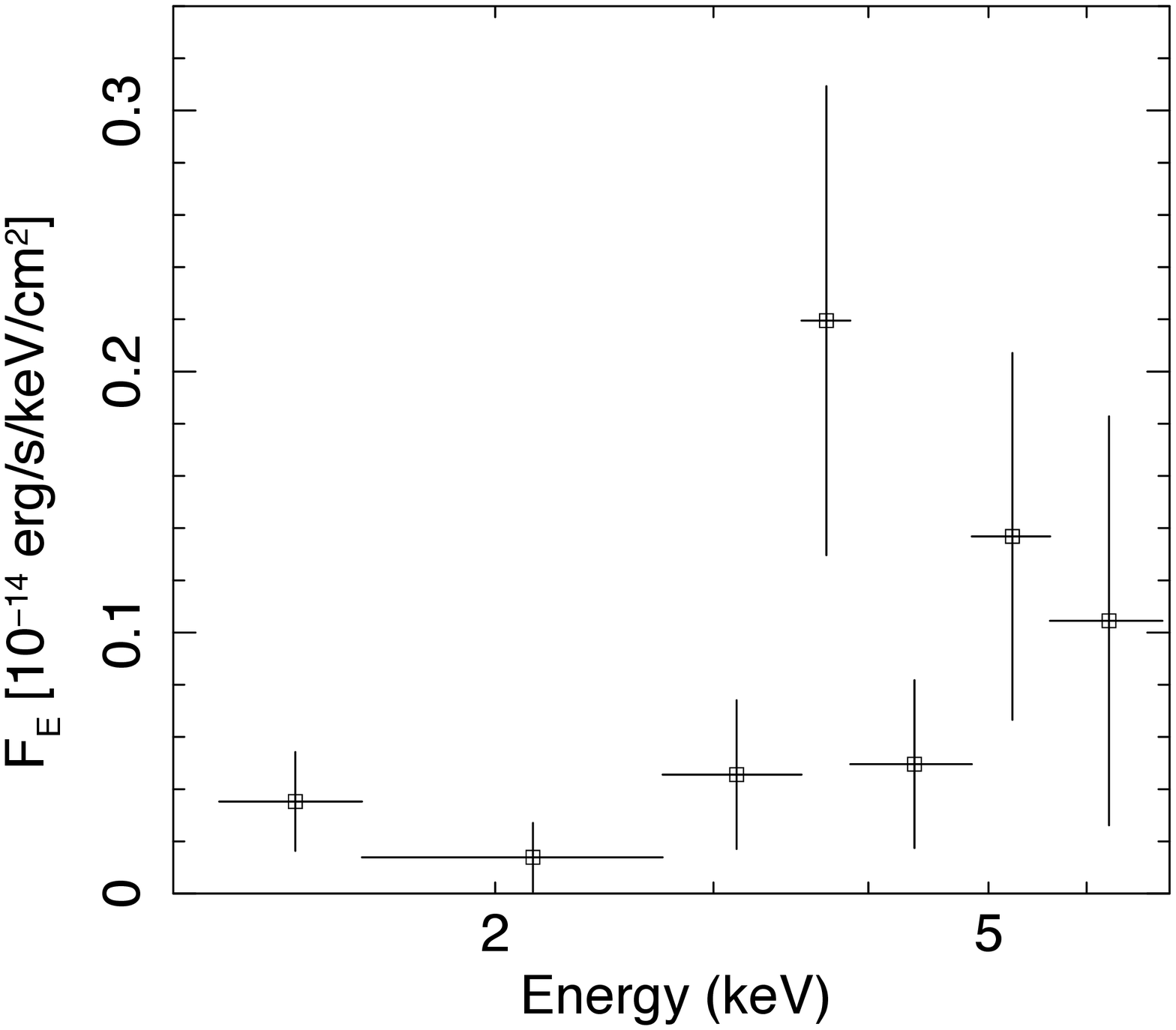}
\caption{Observed-frame \xray\ spectrum of CID1019 (zCOSMOS\_ID=813250) 
at z=0.7302. The flat \xray\ continuum and the tentative iron K$\alpha$ 
emission line at rest-frame energy of $\approx$6.4~keV are indicative of 
strong, likely Compton-thick absorption.}
\label{cid1019_xray}
\end{figure}

Among the analyzed \xray\ spectra, of particular interest is the case of 
source CID1019, corresponding to the zCOSMOS object 813250 at $z=0.7302$ 
($\approx$30 background-subtracted counts). 
A single powerlaw provides $\Gamma\approx-0.5$, which 
is definitely an indication of heavy absorption and is consistent with 
a reflection-dominated spectrum below 10~keV, despite the relatively high 
X/NeV ratio ($\approx40-90$, see Table~\ref{neon23}). 
The tentative presence of an iron K$\alpha$ emission at the rest-frame energy 
of $\approx$6.4~keV with EW$\approx1.7$~keV further supports this scenario 
(see the \xray\ spectrum in Fig.~\ref{cid1019_xray}). 

Eighteen of the 23 Type~2 AGN were detected also by \xmm. Typically, the 
signal-to-noise ratios in \chandra\ spectra are higher than in \xmm\ spectra 
(see \citealt{mainieri2007}) because of the much lower background guaranteed by 
the former satellite. 
Basic \xray\ spectral analysis of \xmm\ data confirms the \chandra\ 
results in terms of spectral properties. Interestingly, there are three sources 
showing \xray\ flux variability by at least a factor of two, with CID1230 
being a factor $\approx$6 fainter in \chandra\ than in \xmm. Unfortunately, 
the limited number of counts ($\approx$30 net counts in the 0.5--7~keV band 
in \chandra, and $\approx$130 in \xmm\ in the same energy band) prevented us 
from investigating the nature of this difference properly.

\begin{table*}
\small
\caption{List of X-ray undetected [NeV]-selected Type~2 AGN}
\label{neon46}
\centering
\begin{tabular}{ccccccr}
\hline
  \multicolumn{1}{c}{zCOSMOS\_ID} &
  \multicolumn{1}{c}{RA$_{\rm \small (J2000)}$} &
  \multicolumn{1}{c}{DEC$_{\rm \small (J2000)}$} &
  \multicolumn{1}{c}{z} &
  \multicolumn{1}{c}{I mag} &
  \multicolumn{1}{c}{T$_{\rm Expo}$} &
  \multicolumn{1}{c}{X/[NeV]} \\
\hline
  \multicolumn{1}{c}{(1)} &
  \multicolumn{1}{c}{(2)} &
  \multicolumn{1}{c}{(3)} &
  \multicolumn{1}{c}{(4)} &
  \multicolumn{1}{c}{(5)} &
  \multicolumn{1}{c}{(6)} &
  \multicolumn{1}{c}{(7)} \\
\hline \hline
  804431 & 150.383041 & 1.745181 & 0.7020 & 22.06 & {\phn}73.97 & $<$81.0\\
  805117 & 150.220490 & 1.729720 & 0.9999 & 22.36 & {\phn}85.50 & $<$48.6\\
  811284 & 150.269028 & 1.891863 & 0.9558 & 22.35 & 191.74      & $<$12.9\\
  811645 & 150.195160 & 1.841540 & 0.8092 & 21.89 & 159.36      & $<$29.7\\
  811887 & 150.144485 & 1.853603 & 0.7297 & 22.25 & 158.61      & $<$10.9\\
  812193 & 150.080215 & 1.849570 & 0.8980 & 21.63 & 159.35      & $<$11.6\\
  812432 & 150.025848 & 1.926400 & 0.6611 & 20.88 & 160.28      & $<$10.0\\
  812665 & 149.971375 & 1.885972 & 0.7301 & 20.88 & 147.61      & $<$37.0\\
  812953 & 149.900970 & 1.947447 & 0.7742 & 21.45 & 166.61      & $<$12.3\\
  813366 & 149.803879 & 1.795451 & 0.6685 & 20.05 & {\phn}78.14 & $<$15.1\\
  813460 & 149.780212 & 1.826555 & 0.6646 & 22.07 & {\phn}79.85 & $<$140.1\\
  813850 & 149.697205 & 1.905195 & 0.6606 & 20.33 & {\phn}76.58 & $<$118.3\\ 
  814229 & 149.607040 & 1.870499 & 0.7618 & 22.47 & {\phn}47.41 & $<$7.9\\
  817886 & 150.244629 & 2.008821 & 0.9603 & 21.74 & 190.76      & $<$119.9\\ 
  818408 & 150.135941 & 2.120217 & 0.6688 & 20.85 & 222.98      & $<$36.5\\
  818478 & 150.122849 & 2.085841 & 0.8946 & 22.29 & 232.48      & $<$48.5\\
  819116 & 149.972672 & 2.049503 & 0.7150 & 22.00 & 174.97      & $<$4.6\\
  819306 & 149.929489 & 2.110626 & 0.9394 & 21.89 & 164.92      & $<$37.6\\
  820589 & 149.606232 & 2.062873 & 0.8796 & 20.78 & {\phn}71.91 & $<$38.9\\
  823097 & 150.469711 & 2.231583 & 0.8044 & 22.23 & {\phn}95.79 & $<$41.8\\
  823162 & 150.458023 & 2.260842 & 0.8489 & 22.07 & 119.31      & $<$20.4\\
  823537 & 150.380005 & 2.128212 & 0.9226 & 21.98 & 169.38      & $<$15.0\\
  824025 & 150.273239 & 2.273062 & 0.8503 & 21.24 & 161.43      & $<$20.6\\
  824548 & 150.163940 & 2.290950 & 0.7480 & 21.89 & 163.79      & $<$6.6\\
  824736 & 150.123535 & 2.149813 & 1.1427 & 22.23 & 199.43      & $<$15.9\\ 
  825958 & 149.884903 & 2.238505 & 0.7026 & 20.95 & 165.84      & $<$74.1\\
  826023 & 149.872528 & 2.162082 & 0.9508 & 21.99 & 151.25      & $<$27.1\\
  826693 & 149.733231 & 2.132358 & 0.6994 & 22.47 & 155.93      & $<$138.9\\
  826908 & 149.695847 & 2.267107 & 1.0246 & 21.46 & {\phn}60.92 & $<$93.4\\ 
  829551 & 150.525146 & 2.456414 & 0.8927 & 21.81 & {\phn}71.55 & $<$73.9\\
  829938 & 150.446548 & 2.366708 & 0.8821 & 21.40 & 163.52      & $<$28.8\\
  830027 & 150.431183 & 2.359476 & 0.9307 & 22.32 & 149.35      & $<$29.2\\
  831655 & 150.112289 & 2.387432 & 0.7567 & 21.68 & 151.48      & $<$98.4\\
  832252 & 150.013412 & 2.333157 & 0.7878 & 21.90 & 157.53      & $<$20.7\\
  832576 & 149.948395 & 2.449379 & 0.7332 & 21.91 & 194.50      & $<$63.6\\
  832803 & 149.905640 & 2.318273 & 0.9274 & 21.49 & 168.74      & $<$23.4\\
  832907 & 149.883438 & 2.373879 & 0.9608 & 21.65 & 149.44      & $<$25.0\\ 
  836868 & 150.477036 & 2.494094 & 0.6793 & 20.78 & {\phn}74.56 & $<$128.5\\
  837072 & 150.439438 & 2.543515 & 1.1475 & 22.44 & {\phn}55.70 & $<$27.3\\
  837589 & 150.347275 & 2.570119 & 0.9214 & 21.47 & {\phn}69.47 & $<$41.5\\
  839719 & 149.963074 & 2.613115 & 0.8916 & 21.64 & {\phn}62.41 & $<$80.1\\
  840744 & 149.772919 & 2.555765 & 0.7353 & 20.68 & {\phn}79.26 & $<$52.4\\ 
  846342 & 150.041229 & 2.634787 & 0.7353 & 20.67 & {\phn}73.21 & $<$14.0\\
  846946 & 149.921646 & 2.638877 & 0.7379 & 21.44 & {\phn}75.52 & $<$47.2\\
  847446 & 149.815765 & 2.650275 & 1.0273 & 22.24 & {\phn}39.18 & $<$73.2\\
  910023 & 150.491974 & 2.458913 & 0.9802 & 21.18 & {\phn}77.54 & $<$63.9\\
\hline
\end{tabular}
\tablefoot{(1) Identification number in the 20k-zCOSMOS Bright Survey; 
(2) optical right ascension and (3) declination; (4) redshift;  
(5) I-band AB magnitude; 
(6) net (``vignetting-corrected") exposure time in the \chandra\ mosaic (ks); 
(7) X/NeV flux ratio using \xray\ aperture photometry from {\sc Acis Extract} 
to compute the rest-frame 2--10~keV flux (see $\S$\ref{sample_and_xray} for 
details). Only one of the sources reported above, zCOSMOS\_ID=813850, has been 
detected in XMM-COSMOS (XMMID=60494, detected in the 2--8~keV band at the 
$\approx4.8\sigma$ level; see \citealt{cappelluti2009}; \citealt{brusa2010}). 
We note that this source is just below the detection threshold in the 
C-COSMOS source catalog.}
\end{table*}

\subsection{[NeV]-selected Type~2 AGN with no X-ray detection}
\label{section46}
\begin{figure}
\centering
\includegraphics[angle=-90,width=\hsize]{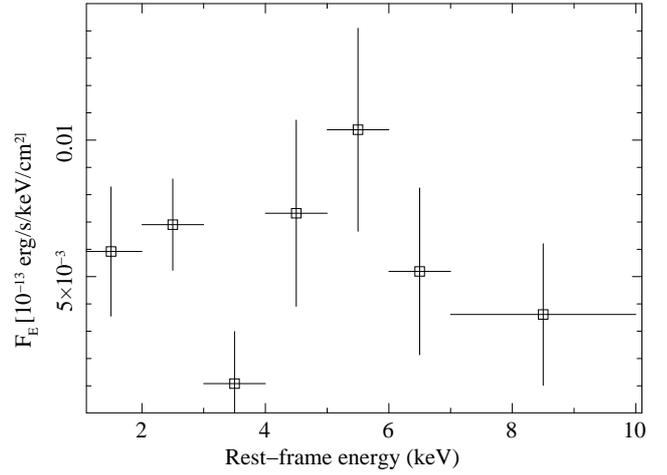}
\caption{Rest-frame 1--10~keV stacked spectrum of the 46 Type~2 AGN 
with no individual \xray\ detection. 
The limited-quality spectrum is flat, suggestive of absorption, but 
apparently no significant iron K$\alpha$ emission line is present 
(EW$<1.5$~keV).}
\label{xrayfitting_n46}
\end{figure}

Results for the sample of 46 \xray\ undetected \nev-selected Type~2 AGN have 
been obtained by \xray\ stacking and X/NeV ratio analyses. 
Because of the limited photon statistics, the former approach is not highly 
effective in constraining the average spectral properties of the sources. 
The fitted photon distribution (130 counts in the observed-frame 
0.5--7~keV band) is consistent with a powerlaw with $\Gamma=0.9\pm{0.5}$, 
indicative of obscuration (see the rest-frame 1--10~keV 
spectrum in Fig.~\ref{xrayfitting_n46}). 
The derived average flux and luminosity of this sample in the 
rest-frame 2--10~keV band are $\approx2.5\times10^{-16}$~\cgs and 
$\approx8.4\times10^{41}$~erg~s$^{-1}$, respectively. 

A step forward in the analysis of this sample consists of placing their X/NeV 
upper limits in Fig.~\ref{xne_histo} (open histogram). 
At face value, 39 of these sources have upper limits below 100, i.e., 
consistent with column densities possibly exceeding 10$^{23}$~cm$^{-2}$ 
(see G10). Nine of these upper limits are located in the region likely 
populated by Compton-thick AGN (i.e., at X/Ne$<$15; G10). 
The average X/NeV ratio for these 46 sources is 9.8 (filled circle in 
Fig.~\ref{xne_histo}). 

To estimate the fraction of Compton-thick AGN in our sample as most reliably 
as possible, we divided the sample of 46 sources into two sub-samples of 
similar size: 22 sources with ``nominal'' X/NeV ratio $<$30 
(hereafter referred to as the ``faint'' sub-sample) and 24 sources with 
``nominal'' X/NeV ratio $>$30 (the ``bright'' sample), 
as also described in M13. Stacked \xray\ photometry for the former 
sub-sample indicates a 1.6$\sigma$ detection in the rest-frame 2--10~keV band, 
compared to the 5.8$\sigma$ detection for the latter sub-sample. 
The derived average X/NeV ratios are 3.6 and 36 for the ``faint'' and 
``bright'' sub-sample, respectively (shown as filled triangle and square in 
Fig.~\ref{xne_histo}, respectively). 

We then stacked the total sample of 46 \xray\ undetected Type~2 AGN, the 
``faint'' and ``bright'' sub-samples; we obtained an average number of 
2.6, 1.1 and 3.9 net counts per source, respectively. 
After verifying that the count distribution of the 46 
sources follows a Poisson distribution (see details in M13), we performed 
10,000 Monte-Carlo runs, each time randomly extracting 46 count values from 
the Poisson distribution and converting these counts into \xray\ fluxes 
(following the prescriptions reported in $\S$\ref{sample_and_xray}). 
Then these \xray\ fluxes were randomly associated to the \nev\ fluxes to 
produce X/NeV flux ratios. As a result of the Monte-Carlo runs, 
an average number of 29.4 sources (2.7 rms) were found at X/NeV$<$15 
and are entitled to be considered reliable Compton-thick AGN candidates, 
bringing the fraction of Compton-thick AGN 
in the present sample from $13\pm5$\%, i.e., 9/69 originally observed at 
X/NeV$<$15, to $43\pm6$\% (i.e., 29.4/69, where the reported error is just 
statistical, and systematic errors related to source selection and stacking 
analysis are likely much higher). 
This fraction compares well with XRB synthesis model predictions 
($\approx50$\%) by \cite{gilli2007}. 

As a consistency check, we ran spectral simulations within {\sc xspec} to 
verify whether the estimated fraction of Compton-thick AGN is consistent 
with the slope of the stacked spectrum of the 46 \xray\ undetected sources 
($\Gamma=0.9\pm0.5$). We assumed that the 30 sources with the most stringent 
limits on the X/NeV ratio are the Compton-thick candidates, whereas the 
remaining 16 objects are Compton thin. We assumed a pure reflection template 
spectrum ({\tt pexrav} in {\sc xspec}) for Compton-thick AGN and an absorbed 
powerlaw for Compton-thin AGN. A soft scattered component was also added in 
both templates, which mainly contributes at energies below 2~keV rest-frame 
(see \citealt{gilli2007} and G10 for details). The \xray\ flux normalization 
of each Compton-thin source was obtained by assuming the real source redshift, 
an intrinsic 2--10~keV rest frame luminosity of 
$L_{2-10 keV}= 400 \times L_{[Ne V]}$, 
where L$_{[NeV]}$ is the measured \nev\ line luminosity, and a column density 
estimated through the $N_H$ vs. X/NeV model relation of G10. Based on the 
measured ratio between the median of the X/NeV upper limits of the sources 
in the ``bright'' sample and the X/NeV value derived from their stacking, we 
assumed that each object has an X/NeV ratio a factor of 2 lower than its 
measured limit. 
A similar procedure was adopted to normalize the spectra of Compton-thick 
candidates, based on the ``faint'' sample. The flux normalization of the 
{\tt pexrav} spectrum was obtained by assuming that it produces $\approx 2\%$ 
of the intrinsic power in the 2--10~keV rest-frame energy band. 
All spectra were simulated by using an average instrumental response for 
C-COSMOS sources and by considering for each source its effective exposure in 
the C-COSMOS mosaic. The 46 spectra were finally co-added using the 
{\tt addspec} ftool. The simulated stacked spectrum contains $\sim110$ photons 
in the 0.5-7 keV band, and a simple powerlaw fit returns a photon index of 
$\Gamma=0.40\pm0.50$. This value is consistent (within 1$\sigma$) with that 
measured in the real stacked spectrum. 
%
%
%
%

\section{The space density of obscured and Compton-thick AGN at $z\approx0.8$}
\label{space_density_obscured}

\begin{figure}
\includegraphics[width=\hsize]{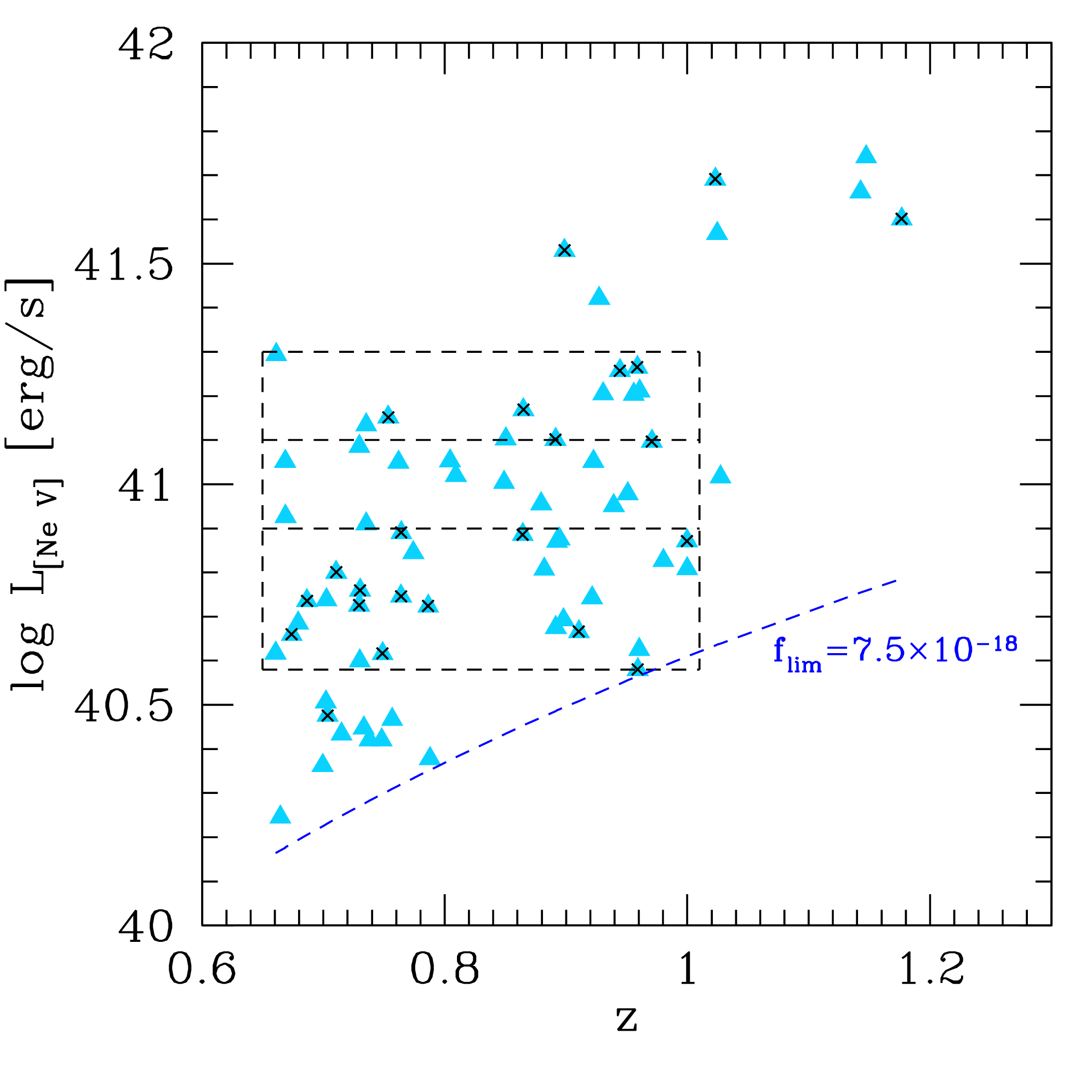}
\caption{Distribution of the 69 \nev-selected Type 2 AGN in the \nev\ line 
luminosity vs. redshift plane. The 23 objects detected in the X-rays are 
marked with a cross. The black dashed lines show the boundaries of the three 
bins used in the computation of the \nev\ luminosity function. 
A total of 50 AGN fall in these three luminosity bins. 
The blue dashed curve shows the 
\nev\ flux limit (in units of \cgs) of the zCOSMOS survey.}
\label{lnev}
\end{figure}

Based on the standard $1/V_{max}$ method \citep{schmidt1968}, we first derived 
the luminosity function of \nev-selected Type~2 AGN at $z\approx0.8$ and then 
the space density of Compton-thick objects among them. 
The distribution of our sources in the \nev\ luminosity vs. redshift 
plane is shown in Fig.~\ref{lnev}. For the luminosity function computation 
we only considered the 50 objects in the redshift range $z=0.65-1.01$ and in 
the luminosity range log$L_{[Ne V]}\approx 40.6-41.3$, which were further 
sub-divided into three luminosity bins (see Fig.~\ref{lnev}). 
We were forced to choose a single, rather broad redshift range to keep a 
reasonable number of objects in each luminosity bin. 
The median redshift of this 50-object sample is $z=0.83$, and does not vary 
significantly across the three luminosity bins. 
The chosen luminosity thresholds guarantee the best completeness level above 
the \nev\ flux limit of the zCOSMOS survey 
($f_{[Ne V]}>7.5\times10^{-18}$\cgs\ at $\approx5\sigma$). 

Sources belonging to our sample were first selected to have $I_{AB}<22.5$ as 
targets for zCOSMOS optical spectroscopy (see \citealt{lilly2009}). 
Therefore, in the $1/V_{max}$ computation, we considered as $V_{max}$ 
the minimum among $V_{[Ne~V]}$, $V_{I_{AB}}$ and $V_{z_{max}}$, where $V_{[Ne~V]}$ 
and $V_{I_{AB}}$ and $V_{z_{max}}$ are the maximum comoving volumes within which 
each object would still be included in the sample based on the \nev\ flux 
cut, $I_{AB}$ magnitude cut, and our adopted redshift cut ($z=1.01$), 
respectively.
In addition, a correction that accounts for both the sampling and success 
rates of zCOSMOS optical spectroscopy needs to be introduced 
(see, e.g., \citealt{bongiorno2010}, B10 hereafter). 
In particular, by defining the target sampling rate (TSR) as the fraction of 
sources observed in the spectroscopic survey compared to the total number of 
objects in the parent photometric catalog, and the spectroscopy success rate 
(SSR) as the fraction of spectroscopically targeted objects for which a secure 
spectroscopic identification was obtained, trends of TSR and SSR as a function 
of object magnitude and redshift were determined 
\citep{zucca2009, bolzonella2010}. A weight can therefore be defined for each 
object in zCOSMOS as $w_i=1/(TSR_i \times SSR_i)$. For reference, in our 
sample $w_i$ ranges from 1.06 to 2.56 and has a mean (median) value 
of 1.87 (1.93). The \nev\ luminosity function in the considered redshift bin 
is therefore given by:

\begin{equation}
\Phi(L) = \frac{1}{\Delta \log L} \sum_i \frac{w_i}{V_{max, i}},
\label{eqphi}
\end{equation}

and the associated statistical uncertainty is given by \citep{marshall1983}: 

\begin{equation}
\sigma_{\phi} = \frac{1}{\Delta \log L} \sqrt{\sum_i \frac{w_i^2}{V_{max, i}^2}}
\end{equation}

For comparison with literature results, the resulting \nev\ luminosity 
function was converted into an \oiii\ luminosity function using the relation 
$L_{[O III]}=9.1\times L_{[Ne V]}$, as derived from objects in our sample that 
are in the redshift range where both lines can be observed. This relation is 
also consistent with the average \nev/\oiii\ ratio derived from the zCOSMOS 
Type~2 AGN composite spectrum (M13) and from other literature 
samples (G10). 
\begin{figure}
\begin{center}
\includegraphics[width=\hsize]{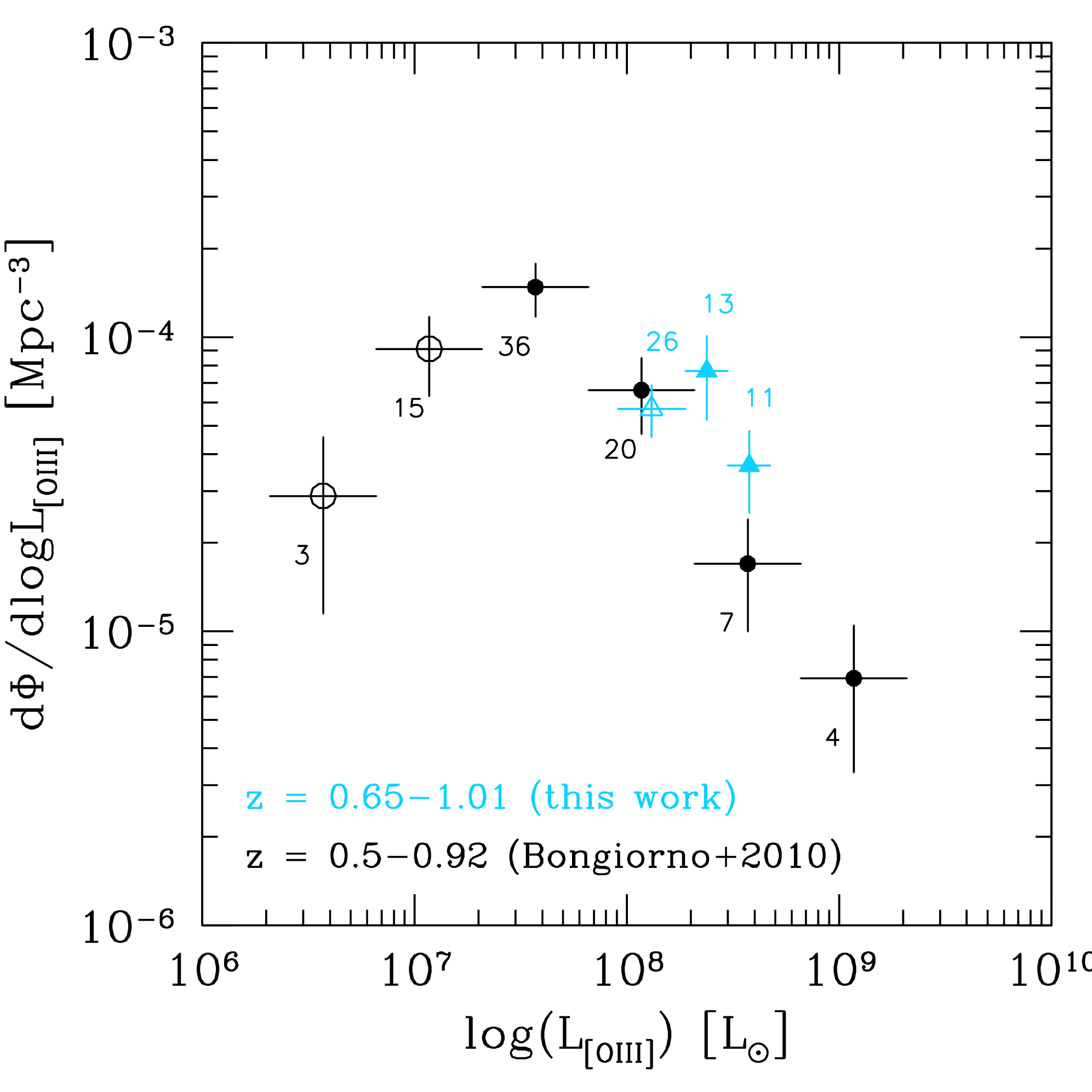}
\caption{\oiii-converted luminosity function of \nev-selected Type 2 AGN at 
$z=0.65-1.01$ in zCOSMOS (light blue triangles; this work) compared with the 
\oiii\ luminosity function of zCOSMOS Type~2 AGN in a similar redshift range 
selected through the ``blue'' line diagnostic diagram (black circles; B10). 
Open symbols highlight those datapoints affected by 
significant incompleteness (see text and B10). 
Small numbers indicate how many objects were used to derive each datapoint.}
\label{lfoiii}
\end{center}
\end{figure}
In Fig.~\ref{lfoiii} we compare the \oiii-converted luminosity function of 
our \nev-selected Type~2 AGN at $z=0.65-1.01$ with that derived by B10 
for zCOSMOS Type~2 AGN selected by means of the 
\oiii/H$\beta$ vs. \oii/H$\beta$ diagnostic diagram \citep{lamareille2004} 
in a similar redshift range ($z=0.50-0.92$). 
With the exception of the datapoint in the faintest luminosity bin, 
which may suffer from some residual incompleteness 
(see below for a more complete discussion), 
the luminosity function of \nev-selected Type~2 AGN at 
$z\approx 0.8$ appears to be a factor of $\sim 2$ higher than that of 
Type~2 AGN in the same redshift and luminosity range derived by B10. 
This confirms the result of M13, who 
showed that a large fraction of Type~2 AGN, selected through their \nev\ 
emission, is in fact missed by the optical classification based on the 
``blue'' \oiii/H$\beta$ vs. \oii/H$\beta$ diagnostic diagram. 
As also discussed by B10 for their \oiii\ luminosity 
function, we note that the drop observed in our luminosity function at the 
faintest luminosities (see Fig.~\ref{lfoiii}) may result from incompleteness 
caused by the two-step selection of our sample. 
In fact, the zCOSMOS magnitude cut at $I=22.5$ may exclude from the 
luminosity function objects that would otherwise be in based on their 
\nev\ flux. We computed the \nev\ luminosity at which the luminosity function 
may suffer from severe incompleteness as follows. 
Based on the Type~2 AGN composite spectrum of M13, we computed the 
3426\AA\ continuum luminosity corresponding to $I_{AB}=22.5$ at $z=1.01$ 
(the upper redshift boundary adopted in the luminosity function computation). 
By combining this value with the \nev\ median EW of the objects included in 
the luminosity function, we derived the line luminosity at which the 
magnitude cut would remove approximately half of the objects. 
This corresponds to $L_{[OIII],half}=1.15\times10^{8}$~\lsun\ and shows that our 
luminosity function is largely incomplete in the lowest luminosity bin. 
%
\begin{figure}
\begin{center}
\includegraphics[width=\hsize,angle=0]{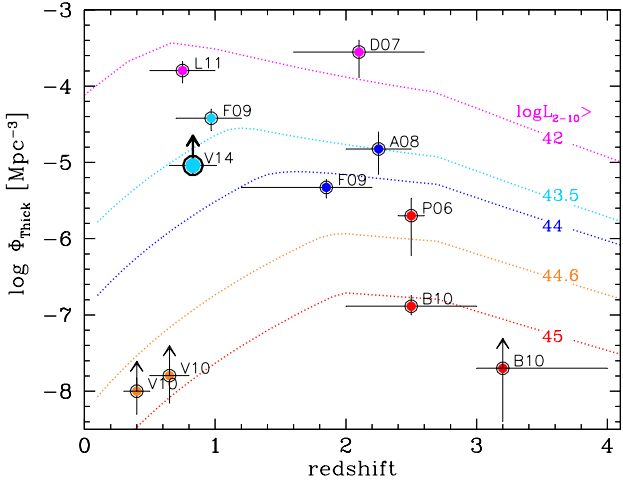}
\caption{Space density of Compton-thick AGN samples with different 
intrinsic 2--10~keV luminosities compared with the XRB model predictions 
\citep{gilli2007}. Datapoints have to be compared with the corresponding 
colour curves; luminosity thresholds are as labeled. References to the 
observed densities, sorted by increasing luminosity threshold, and by 
increasing redshift for the same luminosity threshold, are as follows: 
L11: \cite{luo2011}; 
D07: \cite{daddi2007b}; 
V14: this work; 
F09: \cite{fiore2009}; 
A08: \cite{alexander2008}; 
V10: \cite{vignali2010}; 
P06: \cite{polletta2006};
B10: \cite{bauer2010}.}
\label{xvol_v14}
\end{center}
\end{figure}

Finally, we combined the two datapoints in our luminosity function that are 
most complete (i.e., those at log$L_{[Ne V]}>40.9$ or, equivalently, 
$L_{[O III]}>1.9\times 10^8\ L_{\odot}$) to evaluate the space density of 
Compton-thick AGN in our sample. First we computed the space density of the 
24 AGN in the range $\Delta$log$L_{[Ne V]}=40.9-41.3$ at $z=0.65-1.01$ 
(by using Eq.~\ref{eqphi} and then multiplying back by the 
$\Delta log L$ term), and then we multiplied this density by the fraction of 
Compton-thick objects $f_{Thick}$ as derived in Sect.~\ref{section46} 
(i.e., $f_{Thick}$=0.43). The resulting space density of Compton-thick AGN at 
$z=0.83$ as derived from our sample is then 
$\Phi_{Thick} = (9.1\pm2.1) \times 10^{-6} $ Mpc$^{-3}$. 
To compare this result with the predictions from AGN synthesis models of the 
XRB and literature results, we converted the \nev\ luminosity threshold 
of log$L_{Ne V}>40.9$ into an \xray\ luminosity threshold. In particular, 
since we are interested in the intrinsic AGN power, we used the relation 
log$L_{2-10 keV}$ = 2.6 + log$L_{[Ne V]}$ as derived by G10 
for $unobscured$ AGN, where $L_{2-10 keV}$ is the intrinsic, rest-frame 
2--10~keV luminosity. The derived space density of Compton-thick AGN at 
$z=0.83$ with log$L_{2-10 keV}>43.5$ is shown in Fig.~\ref{xvol_v14} along 
with some measurements of the density of Compton-thick AGN in different 
luminosity and redshift ranges reported by other works and with the 
expectations from the XRB model of \cite{gilli2007} (see also 
\citealt{alexander2011} for a further estimate at $z\approx2$, and 
\citealt{dellaceca2008} and \citealt{severgnini2012} for measurements 
at $z<0.06$). 
Our measurement is in good agreement with both the model 
expectations for log$L_{2-10 keV}>43.5$ and the space density measured by 
\cite{fiore2009} for objects in a similar redshift and luminosity range 
selected through their mid-IR excess in their SED. Since, as discussed earlier, 
even a mild extinction in the NLR can suppress \nev\ emission, 
we regard our selection technique for Compton-thick AGN as clean but 
not complete (see the extended discussion on this issue in 
M13). This is why we plot our measured space density with an 
upward arrow in Fig.~\ref{xvol_v14} (see also \citealt{vignali2010} who 
discuss a similar issue on the space density of \oiii-selected Type~2 AGN 
at $z\approx0.3-0.8$).

\section{Conclusions}
\label{conclusions}
According to XRB synthesis models, a significant fraction of the XRB 
emission at 20~keV (the so-called ``missing XRB'') is produced by Seyfert-like 
Compton-thick objects at $z\approx0.5-1.0$ and with 2--10~keV intrinsic 
luminosities $<10^{44}$~\lum\ (see Fig.~2 of \citealt{gilli2013}). 
To this goal, we have selected a sample of 69 Type~2 AGN drawn from the 
zCOSMOS-Bright Survey on the basis of the presence of the \nev3426\AA\ 
emission line at z=0.65--1.20, a reliable tracer of AGN accretion. 
In this paper we have presented the \xray\ properties of these sources 
(through individual analyses and stacking), and estimated the fraction of 
Compton-thick AGN on the basis of the X/NeV ratio (see G10). 
Finally, we have computed the space density of Compton-thick AGN at 
$z\approx0.8$ and compared it with previous results and \cite{gilli2007} 
XRB model expectations. The main results are as follows: 
\begin{enumerate}
\item[$\bullet$]
Twenty-three of the 69 \nev-selected Type~2 AGN were detected in the 
\chandra-COSMOS observations. Their spectral properties suggest a wide 
range of absorption, going from limited obscuration for few sources among 
those with most \xray\ photons to heavy obscuration, although formally none 
of these objects has X/NeV ratio falling in the Compton-thick AGN region. 
\item[$\bullet$]
Using \xray\ stacking analysis for the individually \xray\ undetected 
Type~2 AGN coupled with Monte-Carlo simulations provides a fraction of 
$\approx$~43\% of Compton-thick AGN in the sample of 69 \nev-selected 
Type~2 AGN. 
\item[$\bullet$]
The derived space density of Compton-thick AGN at $z=0.83$ is 
$\Phi_{Thick} = (9.1\pm2.1) \times 10^{-6} $ Mpc$^{-3}$. This value is in 
good agreement with XRB model expectations for log$L_{2-10 keV}>43.5$ 
AGN and with the previously measured space density for Compton-thick objects 
in a similar redshift and luminosity range \citep{fiore2009}. 
\item[$\bullet$]
The \oiii\ luminosity function, ``converted'' from the measured \nev\ 
luminosity function, is a factor $\approx2$ above that computed by B10 
for Type~2 AGN selected in the zCOSMOS survey at similar 
redshifts using the \oiii/H$\beta$ vs. \oii/H$\beta$ diagnostic diagram. 
This result indicates that any optical selection method based on spectroscopy 
and diagnostic diagrams is not complete in finding obscured AGN. 
Therefore, to search for the most heavily obscured objects, a ``panchromatic'' 
observing strategy is the way to go (e.g., Delvecchio et al., in preparation). 
\end{enumerate}

\begin{acknowledgements}
CV thanks M. Bolzonella for useful discussion. 
Financial contribution from ``PRIN--INAF 2011" and ``PRIN--INAF 2012" 
is acknowledged. KI thanks support from Spanish Ministerio de C\'\i{}encia 
e Innovac\'\i{}on (MICINN) through the grant (AYA2010-21782-C03-01).
\end{acknowledgements}


\end{document}